\documentclass{aa}  
\usepackage{graphicx}
\usepackage{txfonts}
\usepackage[usenames]{color}
\usepackage{ulem}

\begin{document}

   \title{Twenty years of PWV measurements in the Chajnantor Area}

   \subtitle{}

   \author{F. Cort\'es  \inst{1}, K. Cort\'es  \inst{1}, R. Reeves  \inst{1}, R. Bustos  \inst{2}  \and S. Radford \inst{3}}

   \institute{ CePIA, Departamento de Astronom\'ia, Universidad de Concepci\'on, Casilla 160 C, Concepci\'on, Chile \\      
                \email{fercortes@udec.cl}
             \and
            Laboratorio de Astro-Ingenier\'ia y Microondas, Facultad de Ingenier\'ia, Universidad Cat\'olica de la Sant\'isima Concepci\'on, Alonso de Ribera 2850, Concepci\'on, Chile
             \and
             Smithsonian Astrophysical Observatory, Hilo, Hawaii.     
                }

   \date{ Accepted June 9, 2020}

 \abstract 
 {Interest in the use of the Chajnantor area for millimeter and submillimeter astronomy is increasing because of its excellent atmospheric conditions. Knowing the general site annual variability in precipitable water vapor (PWV) can contribute to the planning {of new observatories in the area}.} {We seek to create a $20$-year atmospheric database ($1997-2017$) for the Chajnantor area in northern Chile using a single common physical unit, PWV. We plan to extract weather relations between the Chajnantor Plateau and the summit of Cerro Chajnantor to evaluate potential sensitivity improvements for telescopes fielded in the higher site. We aim to validate the use of submillimeter tippers to be used at other sites and use the PWV database to detect a potential signature for local climate change over $20$ years.} {We revised our method to convert from submillimeter tipper opacity to PWV. We now include the ground temperature as an input parameter to the conversion scheme and, therefore, achieve a higher conversion accuracy.}{We found a decrease in the measured PWV at the summit of Cerro Chajnantor with respect to the plateau of $28\%$. In addition, we found a PWV difference of $1.9\%$
with only $27$ m of altitude difference between two sites in the Chajnantor Plateau: the Atacama  Pathfinder  Experiment (APEX) and the Cosmic Background Imager (CBI) near the  Atacama  Large  Millimeter  Array (ALMA) center. This difference is possibly due to local topographic conditions that favor the discrepancy in PWV. The scale height for the plateau was extracted from the measurements of the plateau and the Cerro Chajnantor summit, giving a value of $1537$ m. Considering the results obtained in this work from the long-term study, we do not see evidence of PWV trends in the $20$-year period of the analysis that would suggest climate change in such a timescale.}{}


\authorrunning{Cortés, F. et al.}
   \maketitle

\section{Introduction}

The Chajnantor area, a high altitude site in the Chilean Altiplano, is recognized as one of the best sites in the world for the millimeter, submillimeter, and mid-infrared astronomical observations \citep{bustos14}. Located at $23^{\circ}$00'54"S $+67^{\circ}$45'45"W and 5080 m in altitude, this area has averages of 555 mbar in atmospheric pressure and 273 K in temperature. Cerro Chajnantor, located next to the plateau, has a summit altitude of 5640 m and has averages of 518 mbar in atmospheric pressure and 268 K in temperature. The combination of high altitude and extreme dryness makes the area one of the most accessible sites for submillimeter astronomy in the world. 

The outstanding observing conditions at Chajnantor have prompted the installation of world-class astronomical observatories. These include the Atacama Large Millimeter Array (ALMA) \citep{wotten09}, the Atacama Pathfinder Experiment (APEX) \citep{gusten06}, the Atacama Cosmology Telescope (ACT) \citep{kosowsky03}, the Cosmic Background Imager (CBI) \citep{padin02}, and future experiments such as the Tokyo Atacama Observatory (TAO) \citep{motohara11}, Simons Observatory (SO) \citep{ade19}, CCAT-prime \citep{ccat-paper}, and the Leigthon Chajnantor Telescope (LCT)\footnote{http://www.astro.caltech.edu/twiki\_cstc/bin/view}.

In the context of atmospheric characterization of the site, the first measurements of observing conditions on the Chajnantor Plateau began in April 1995 \citep{radford98}. Since then, many atmospheric characterization instruments have been deployed to the site. These instruments have covered multiple time epochs and geographical locations, depending on the specific interests of each research group or astronomical observatories.

There are a number of publications in the literature referring to measurements of precipitable water vapor (PWV) and weather conditions at the Chajnantor area and their implications for astronomical observations. As an example of these studies, based on data from radiosonde launched from the Chajnantor Plateau, \cite{giovanelli01} suggested peaks such as Cerro Chajnantor would be drier than the Chajnantor Plateau. Observations at terahertz frequencies from Sairecabur (5500\,m) indicated this was correct \citep{marrone04, marrone05}; in addition, simultaneous $350$ $\mu m$ measurements from the Chajnantor Plateau and Cerro Chajnantor provided confirmation of these observations. \cite{delgado99, radford08} derived a conversion scheme to obtain PWV from radiometric measurements of the $183$ GHz water emission line, while longer-term climatology studies in the Chajnantor area were presented in \cite{bustos00, otarola05, otarola19}. Estimates and forecasts of PWV\ for the Chajnantor area using the Geostationary Operational Environmental Satellite (GOES) were presented in \cite{marin15}, and in addition, seasonal and intraseasonal variations for PWV were studied in \cite{marin17}. In \cite{holdaway04}, the relationship between global warming and PWV for the area was studied in a period of six years with inconclusive results. An atmospheric measurement campaign was performed on Cerro Toco, , a site located within the Chajnantor area, by \cite{turner10} in 2009 and their data were used in \cite{cortes16} to assess the vertical distribution of PWV in the area. The atmospheric transparency was studied in depth, spanning long periods of time by \cite{radford00, giovanelli01, radford11, radford16}. In addition, PWV ratios between the plateau and Cerro Chajnantor were presented in \cite{bustos14} for a time span of five days and expanded over longer periods in \cite{cortes16}, including estimates for atmospheric scale heights. 

Studying the dynamics of the atmospheric conditions is crucial for short wavelength radio astronomy, since it directly affects data quality in the form of attenuation and differential phase delay of the astronomical signals as they reach the telescopes. The timescale for these effects is very broad and includes intra-day variations, seasonal patterns, and even exhibits long-term features such as the El Ni\~no--Southern Oscillation (ENSO), as an example. On short timescales, of order seconds, atmospheric variability can affect radio seeing on large aperture radio telescopes through anomalous refraction \citep{olmi01}. This paper is a revision and refinement of a previous study of the mentioned atmospheric dynamics and tropospheric distribution of the water vapor over the Chajnantor area. To achieve this goal and to improve the quality of our data, we revisited the methodology for the conversion of the submillimeter tipping radiometer (tipper) data into PWV presented in \cite{cortes16}. As explained in the paper, the conversion method now includes information on the surface temperature at the instrument location at the time of the measurement, which reduces the uncertainty in the results. The new scheme is applied to all the tipper data that exists for the area, and we achieve an expansion of our local atmospheric database now spanning from 1997 to 2017. We validate the results of the conversion scheme by comparing these findings with data from other instruments fielded in the area based on various technologies. We demonstrate that the tippers do not need another instrument as a reference for absolute PWV calibration, as their PWV converted results are only a few percent off the data from the considered measurement standards. This is relevant because the tipping instruments can now be treated as independent integrated PWV measurement devices if they are coupled with an appropriate atmospheric model for the site under study, as discussed in \cite{cortes16}. Ultimately, we hope that existing and future observatories will find this information useful in the planning of their logistical activities at the site, therefore maximizing their scientific return.

\section{Instruments and software tools}

In this section, we provide details on the instruments that were used for this study. The data used in this work were obtained with $183$ GHz radiometers and with $350$ $\mu$m tipping radiometers. These are different in their technical characteristics and measurement techniques, location, altitude, and deployment time coverage. A summary of these instruments including specific location, altitude, and nomenclature is presented in Table \ref{tab-instr}; more details about these instruments are shown in Table 1 \citep{cortes16} and in citations therein. The following are noteworthy comments about these measuring devices:

\begin{table}[h]
\caption{\label{t7}Instruments used in this study and their location, altitude, and ID for reference in this paper.}
\centering
\begin{tabular}{lcccc}
\hline\hline 

Instrument       & Location      &    Altitude (m)  &Time & ID \\ 
                        &                    &                         & span&  \\ \hline \hline   \\
 APEX              &  Chajnantor  &  5107              & 2006  & APEX \\ 
radiometer       &  Plateau        &                        & to 2017    &\\  \\ \hline  \\

                         &  Chajnantor        &  5080   &  1997 &TA-1 \\ 
                         &  Plateau             &             &   to 2005 & \\ 
                         &     (NRAO)         &             &    \\ \\
                     
 Tipper              & Chajnantor  & 5080     &2005 &TA-2 \\ 
 radiometer A    &  Plateau       &             & to 2010 &\\ 
                         & (CBI)            &             & \\ \\

                         & Chajnantor  &  5107  &2011&  TA-3\\ 
                         &  Plateau       &           & to now& \\  
                         &   (APEX)     &           &  \\ \\  \hline \\

                         &  Chajnantor  &  5080    &2000     & TB-1  \\ 
                         &  Plateau       &              & to 2005 &\\ 
                         &   (NRAO)     &              &  \\ \\

Tipper               & Chajnantor  & 5080   & 2005 &TB-2 \\ 
 radiometer B    &  Plateau      &            & to 2009&  \\ 
                          &  (CBI)         &     & \\ \\

                         &  Cerro          &  5612   & 2009 &TB-3 \\ 
                         & Chajnantor   &            &    to now      & \\   \\ \hline  \\
                         
WVR UDEC      & Cerro           & 5612   & 2011  & UdeC \\   
                          &  Chajnantor  &         & to now&\\

\hline
\end{tabular}
\label{tab-instr}
\end{table}

Water vapor radiometers (WVR): The APEX and Universidad de Concepción (UdeC) WVRs provide measurements of the atmospheric brightness temperature of the $183$ GHz water line and over a number of defined bandpasses to spectrally characterize the emission. These instruments are based on Schottky-diode mixers coupled with a baseband analog filter bank \citep{delgado99}. Because these systems operate at room temperature, the receiver noise temperature for these instruments is usually above $1500$ K. Once the spectral data are taken, an atmospheric model is typically used to fit the observations and estimate the PWV. Data from APEX WVR were obtained on the APEX webpage\footnote{http://www.apex-telescope.org/}.

Tipping radiometers of $350$ $\mu$m: These instruments measure the sky brightness at different air masses in order to estimate the atmospheric opacity. They are built with bandpass filtered pyroelectric detectors and integrate the incoming radiation over a $103$ GHz band, centered at $850$ GHz. The calibration uses measurements of absorbers at known physical temperatures to determine the antenna temperature at each air mass. The measurements are fitted to an exponential function to estimate the radiometric temperature of the  sky and the atmospheric opacity \citep{radford16}. The data from tipping radiometers and UdeC WVR are available online\footnote{http://doi.org/10.5281/zenodo.3880373}.

The computational tools used in this work were Atmospheric Model 9.0 (AM) \citep{paine16}, Python version 2.7.2, and Tool for Operations on Catalogues And Tables (TOPCAT) version 4.1. We note that the atmospheric transmission at microwaves (ATM) model has been extensively used in atmospheric monitoring at observatories, such as APEX and ALMA. We compared the AM and ATM models and their differences are $<3\%$ over the bands of interest.

\section{Improved conversion from 350 $\mu$m tipper opacity to PWV}
A major goal of our research is to bring all the existing atmospheric database for the Chajnantor area to a common physical unit, PWV. This, in turn, has the aim of understanding the distribution of the water vapor over the area along with its temporal variability. With this study we are also attempting an insight into climate variations or trends, taking advantage of the more than $20$ years of data collected at the site.

\begin{figure}[h]
\resizebox{\hsize}{!}{\includegraphics{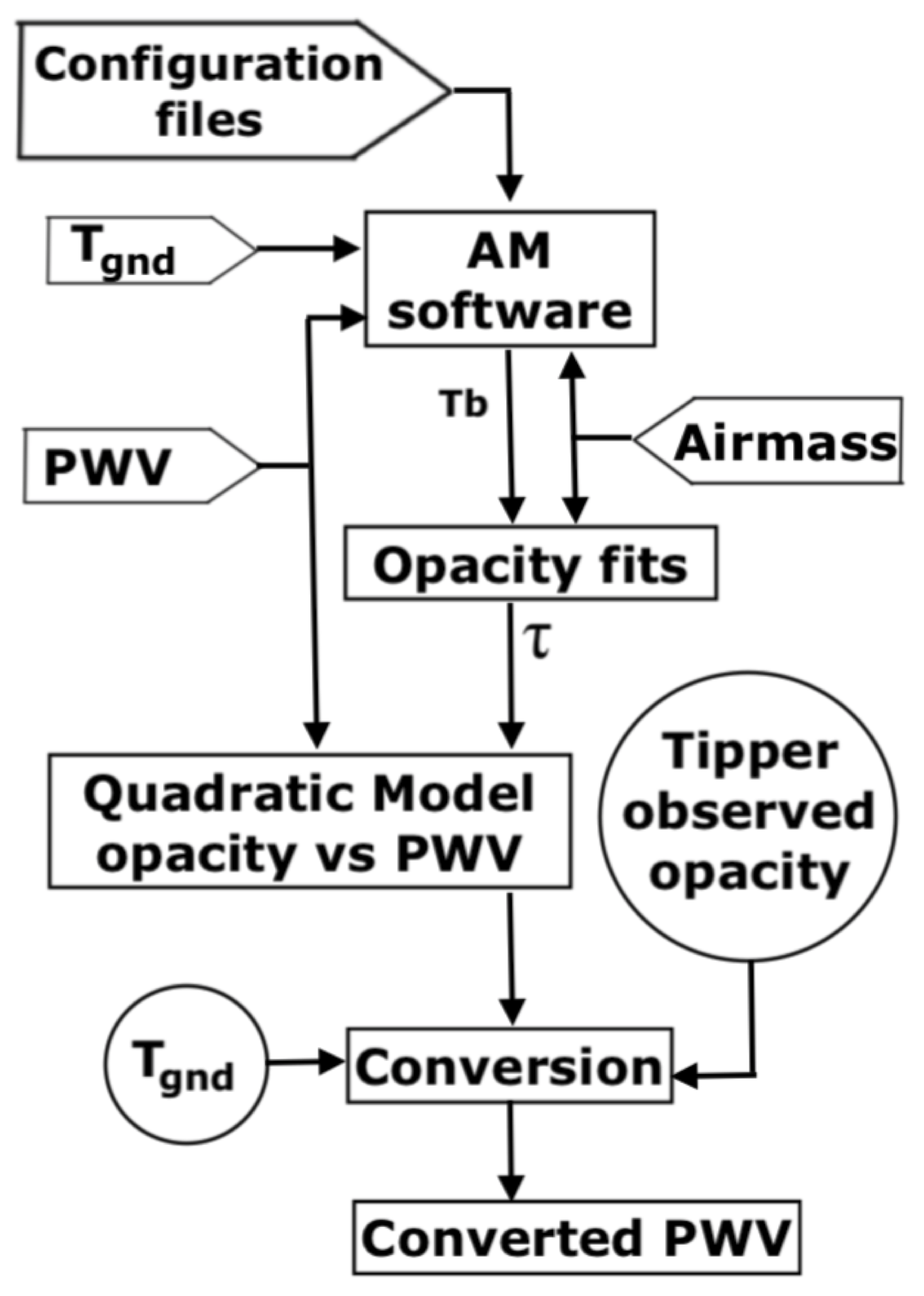}}
\caption{Illustrative block diagram describing the procedure to convert from measured tipper opacity to PWV. The one-end pointed rectangle blocks are considered inputs to the procedure, while the full rectangles are procedural actions. The circles represent experimental data coming from the submillimeter tipper.}
\label{figure110}
\end{figure}

The WVR mounted on the APEX telescope is considered a comparison standard for PWV measurements at Chajnantor and has been operational since 2006. However, the measurements carried out by the 350 $\mu$m tipper radiometers long predate those from APEX, spanning from 1997, and can be used to establish a longer time baseline. The submillimeter tipper delivers atmospheric opacity in Neper units [$N_p$] and a method to convert from tipper opacity to PWV in millimeters of integrated columns was presented in \citep{cortes16}. The method models the mechanical and radiometric functionality of the instrument and a conversion
from measured opacity to PWV was devised using the AM software to solve for the atmospheric radiative emission. A basic block diagram describing the procedure to convert from tipper opacity to PWV is shown in Figure \ref{figure110}.

\begin{figure}[h]
\resizebox{\hsize}{!}{\includegraphics{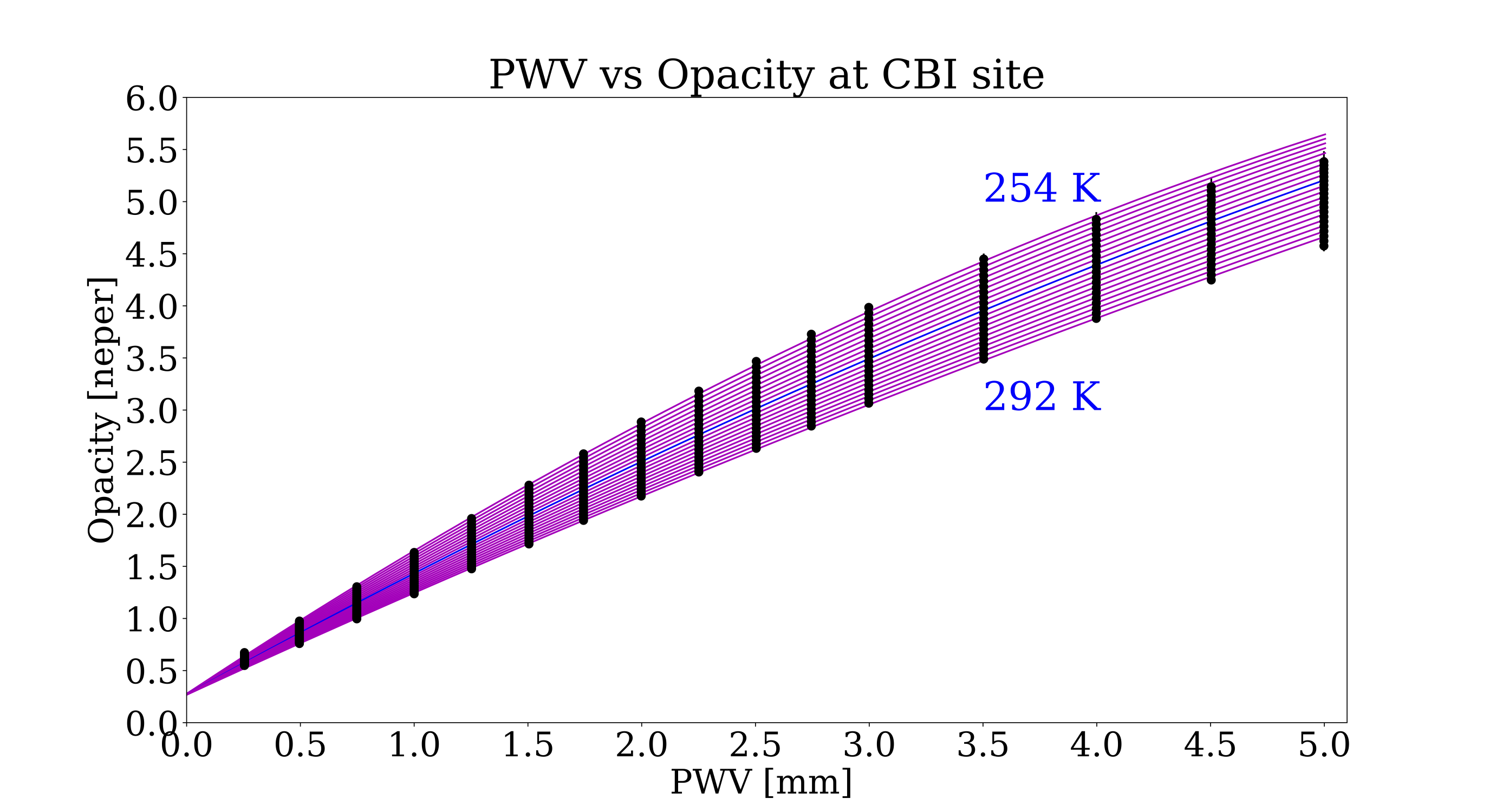}}\\ \resizebox{\hsize}{!}{\includegraphics{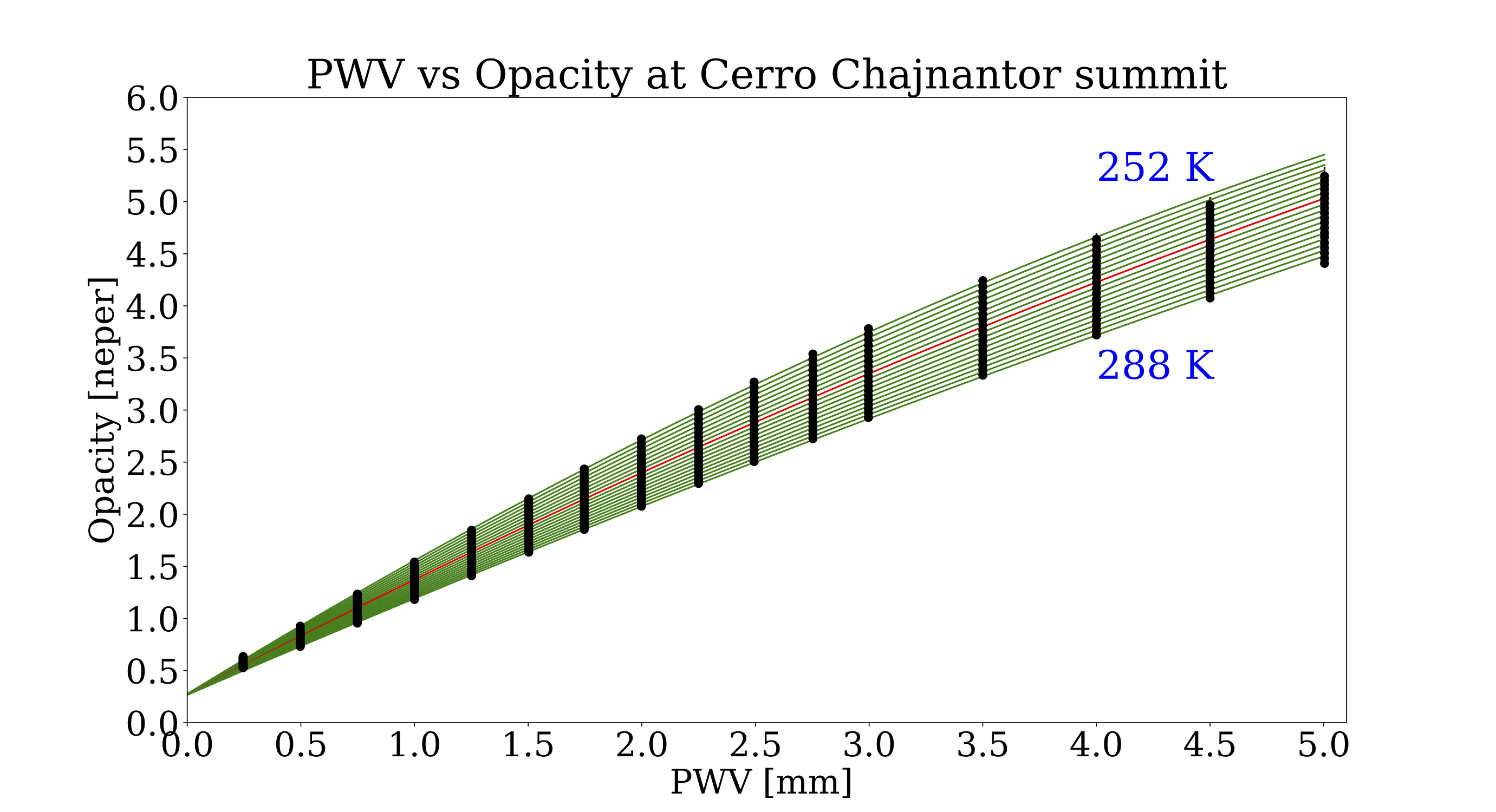}}
\caption{Tipper opacity simulations. The black dots indicate values extracted from the AM simulations. The lines correspond to weighted quadratic fits to the simulated data and for each $T_{gnd}$. The upper panel presents the situation for the Chajnantor Plateau, while the lower panel shows the data and fits for the summit of Cerro Chajnantor. The blue (red) line in the middle of the distribution corresponds to the results for the average ground temperature at the plateau (Cerro Chajnantor) of $272.4$ K ($268.6$ K), as previously used in \citep{cortes16}.}
\label{figure3}
\end{figure}

In the first version of the tipper opacity to PWV conversion method \citep{cortes16}, the ground temperature ($T_{gnd}$) input to the AM model was set to a constant value, which was taken as the historical average ground temperature for each site that was analyzed. However, $T_{gnd}$ is considered as the tropospheric end of the AM layered radiative transfer configuration that models the atmosphere and, therefore, it has significant impact on the simulation results and modeling of the atmospheric conditions. We know from local weather data that $T_{gnd}$ varies from $-23^\circ$C to $+14.5^\circ$C (APEX public database) in the Chajnantor area, and not taking the diurnal oscillation into account introduces unnecessary uncertainties in the conversion process. Following \cite{cortes17}, we started using the actual tropospheric temperature at the instrument location as a variable input to the AM model runs. The temperature values used were measured at each site by local weather stations, concurrently with the opacity measurements available for conversion.   
The method to convert from $350\>\mu m$ tipper opacity to PWV includes the components explained in~\cite{cortes16} and adds new procedures, which are detailed below. The original components can be summarized as follows: a) A multilayered atmospheric model for each site is created, which is an input configuration file for the AM software, and is entirely based on National Aeronautics and Space Administration, Modern-Era Retrospective analysis for Research and Applications, Version 2 (NASA MERRA-2) reanalysis data \citep{molod15}. b) The use of the AM software simulates the mechanical and radiometric operation of the tipper, which gives an effective temperature for the atmosphere averaged over the 350 $\mu$m band for a given value of PWV,  and is weighted by the spectral response of the filter located at the input of the tipper. c) We extract the opacity and atmospheric brightness temperature by fitting an exponential function to the simulated temperature versus tipper zenith angle. From now on, we detail the new procedures of the proposed methodology with the aim of reducing the uncertainty in the conversion to PWV and obtaining an accurate database available for climate studies:

d) Opacity to PWV conversion: This step includes a relevant difference from our previous work \citep{cortes16}, in that the relationship between opacity and PWV was presented as a linear characteristic and the average $T_{gnd}$ was used as a fixed value input to the simulation. Introducing $T_{gnd}$ as a swept variable input and extending the PWV range input to the simulation, we found a quadratic relationship between the PWV and tipper opacity, as shown in Figure \ref{figure3}. In the upper panel of the figure, the conversion between the PWV and tipper opacity for the Chajnantor Plateau is presented, while the lower panel corresponds to the modeled conversion for the summit of Cerro Chajnantor. For both plots, there is an opacity pedestal at zero PWV, which is known as the dry opacity component that includes other atmospheric gases, such as oxygen and ozone. We note that the estimated dry opacity from the simulations for the plateau and the Cerro Chajnantor are 0.272 Np and 0.271 Np, respectively. The impact of $T_{gnd}$ in the emissivity of the water molecule was recently noted and discussed in~\cite{radford16}, in agreement with our analysis and supporting the changes in our conversion scheme for the submillimeter tippers.

The 350 $\mu$m tipper opacity is measured at a certain ground temperature. We use that specific ground temperature to derive the conversion to PWV, as depicted in~Figure \ref{figure3}. The conversion to PWV is derived by interpolating the model grid points in~Figure \ref{figure3} using the actual $T_{gnd}$ and the measured tipper opacity.

e) WVR sampling normalization: This is another important difference from our previous work. The APEX radiometer output rate is one sample per minute, while the tipping radiometer provides a measurement every 13 minutes, which is time stamped at the middle of the measurement cycle. In our previous work \citep{cortes16}, the data from both instruments were matched in time in order to be processed, which induced a delay between the two atmospheric observations and therefore an extra source of uncertainty. In this version of the analysis, an average windowing filter with length of 13 samples was applied to the WVR measurements and only the samples prior to the time stamped data were considered. The outcome is a more precise and representative comparison between the results of both instruments.

\begin{figure}[hbt]
\resizebox{\hsize}{!}{\includegraphics{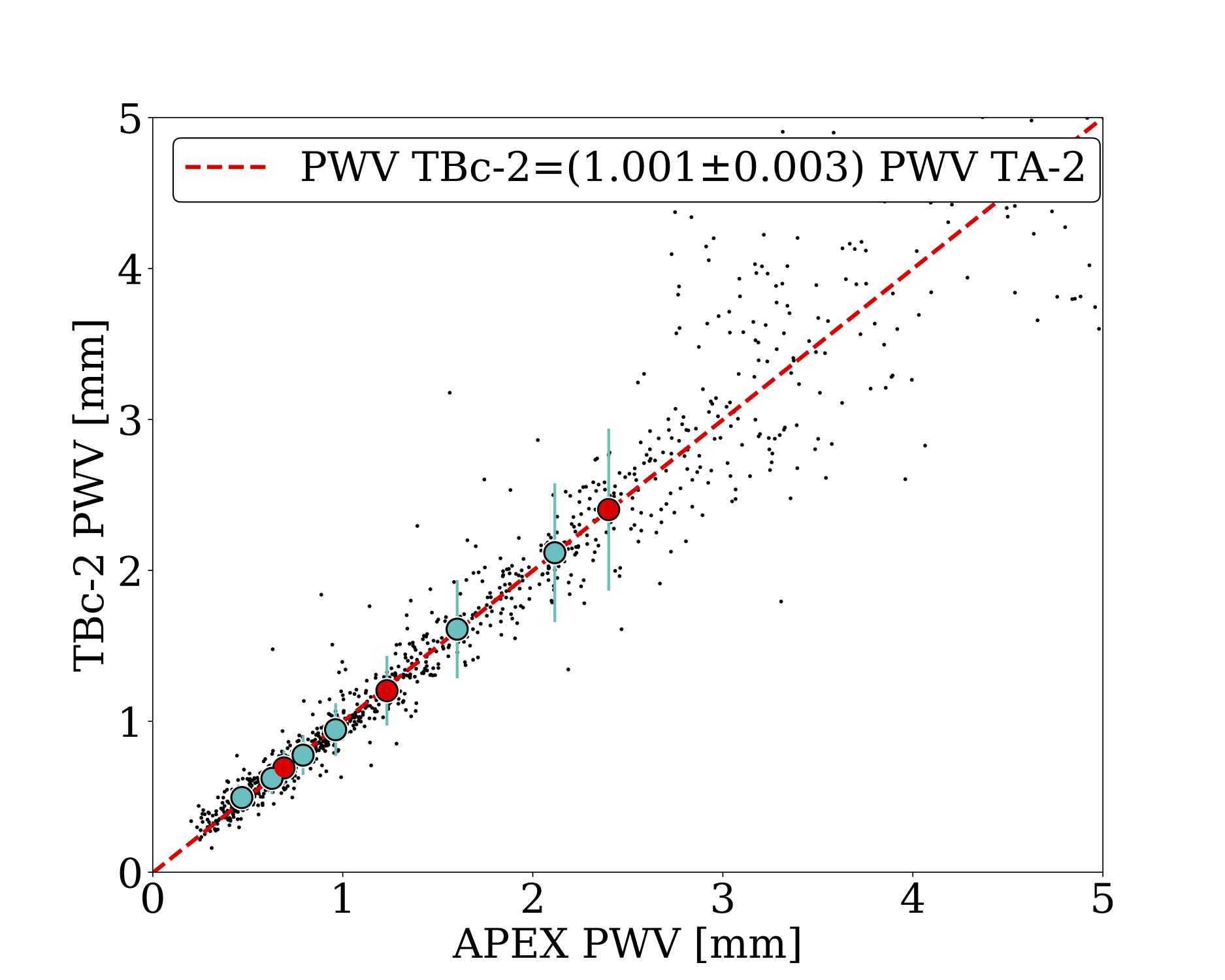}}
\caption{Corrected tipper data for a shared location measurement campaign, for tippers TA and TB, at the Chajnantor Plateau (CBI) site, covering years 2005 and 2006. The cyan circles are the PWV quantiles used in linear regression. The red circles indicate the quartiles ( 25\%, 50\%, and 75\%). Error bars are y-axis standard deviation for each quantile.}

\label{fig8}
\end{figure}

As mentioned above, the PWV output of the tipper data that pass through the conversion procedure is readily available to be used for atmospheric analysis. The only correction that is applied to the PWV data is the cross calibration between both tippers to match their results when these are colocated. Both tippers shared the same site over years $2005$ and $2006$; therefore, this was an opportunity to evaluate for measurement consistency between the tipping instruments. This campaign showed that the tippers were consistent to a figure of $1.12\%$ in PWV, taken from the slope in PWV vs PWV ratio, and that factor was applied to the TA tipper to cross calibrate both data outputs as follows: $PWV_{TB}^{corr} = PWV_{TB}/1.012$. Once this factor was applied, the PWV versus PWV diagram for this concurrent measurement session delivered the results shown in~Figure \ref{fig8}. The slope between both instrument outputs was $1.001 \pm 0.003$, with a Pearson correlation factor of $0.94$.

The error in the slope was calculated using the Equation \ref{eq2}. This equation was used to estimate the standard error of a slope in simple linear regression method \citep{crunk14}, as follows:

\begin{equation}
\hspace{20mm}STD_{slope}\>=\> \displaystyle\frac{\sqrt{\displaystyle\sum\frac{(Y_{model}-Y_{data})^2}{N-2}}}{\displaystyle\sqrt{\sum(x-\bar{x})^2}}
\label{eq2}
,\end{equation}

where $Y_{model}$ is the obtained value predicted by the linear regression, $Y_{data}$ is the real value, $N$ is the number of data involved in the linear regression, x is the real value, and $\bar{x}$ is the mean of x values.

\section{Estimating PWV ratios}
Estimating the PWV ratio between datasets from different sites or instruments is crucial if the aim is to assess the distribution of water vapor on a certain area, compare how much drier a site is than another, or evaluate the scale height for water vapor over a certain site. The method to estimate the PWV ratio we applied so far, detailed in \cite{cortes16}, is to fit a line to a time-matched PWV vervus PWV diagram and extract the slope directly from the fit. It was found that the result in the estimate for the slope is very sensitive to how clumped the data are near the origin of the diagram as well as what exact range is assumed as the linear regime of the system prior to the analysis. The nature of the PWV datasets for the Chajnantor area is such that the data in the PWV versus PWV diagram are mostly clumped near the origin and small variations on the distribution of the data for low values of PWV can have significant effects on the result for the PWV ratio, which may lead to wrong interpretations about the physics governing a certain site. 

In this paper, we applied a variation to the method of estimating the PWV ratio. We instead used the cumulative distribution of PWV for a given site of the data on each axis of the diagram. Each cumulative distribution is divided into $20$ quantiles; therefore, each quantile for the involved sites forms a pair in the PWV versus PWV diagram and is used for parameter extraction with the line regression algorithm of choice. We believe this method uses a more faithful statistical representation of each dataset compared to the previous method, providing a more robust estimate for the sought PWV ratio. The data used for the line regression consider up to $75\%$ of the full dataset because the remaining data are found to be dominated by measurement uncertainty and, for high PWV, the submillimeter tippers respond nonlinearly. The linear regime used for the analysis in PWV has an upper limit of $3$ mm, which is considered to be useful for millimeter and submillimeter radio astronomy interests. Figure~\ref{fig8} shows the results of the new method, in which cyan circles denote some of the $20$ quantiles mentioned before (10\%, 20\%, 30\%, 40\%, 60\%, and 70\% specifically), while the red circles indicate the quartiles, for reference.\\

\section{Validation of submillimeter tipper data from revised conversion procedure}

In our previous work \citep{cortes16}, the PWV measurements from the WVR at APEX are considered the calibration standard for our atmospheric characterization database. In this work, it was found that the submillimeter tipper radiometer measurements require a small correction to match the APEX WVR measurements, $(< 2\%)$ that is, to produce a PWV versus PWV diagram with a slope of unity. Therefore, we can safely consider TA and TB, namely the submillimeter tipper radiometers, as the two instruments providing measurement outputs independent of any external calibrator. The implication of this finding is that the tippers can operate independently, installed at multiple sites of low PWV, and obtain scientifically valid PWV data provided that the suggested conversion method explained in Section 3 is followed. To confirm this statement, in the following we compare the measurements from tippers against other local instruments using PWV versus PWV scatter plots. 

\begin{figure}[h]
\resizebox{\hsize}{!}{\includegraphics{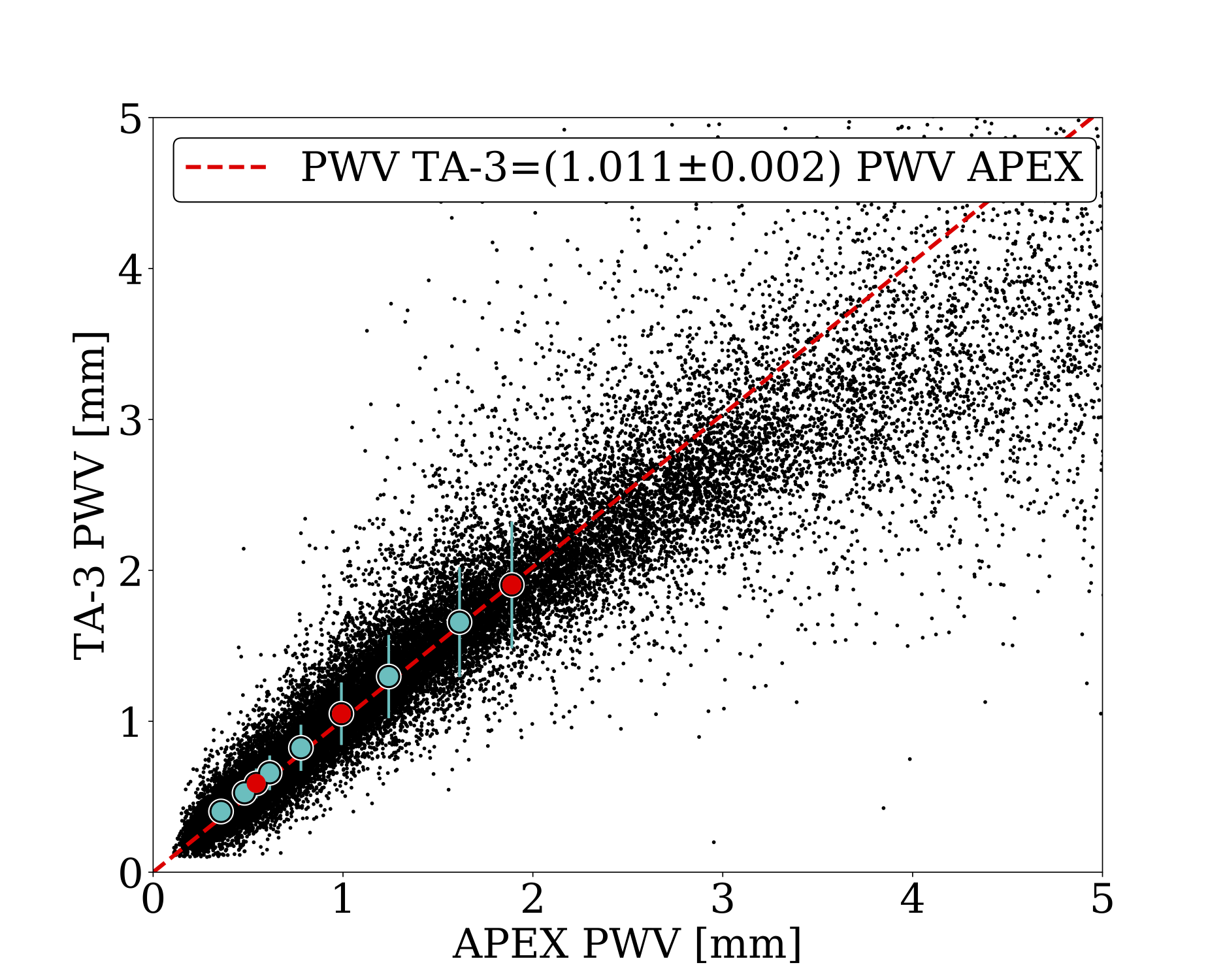}}
\caption{Comparison between the APEX WVR measurements and the PWV from tipper TA-3 for 2011-2014, when both instruments coincided in both site and time. The cyan circles are the PWV quantiles used in linear regression. The red circles indicate the quartiles ( $25\%$, $50\%$, and $75\%$). Error bars are y-axis standard deviation for each quantile. As noted in the text, TA-3 shows a small offset with APEX, which shows the measuring independence of the tipper instrument when the proposed conversion is used.}
\label{figure7}
\end{figure}

The tipper TA was colocated with the APEX WVR for a span of three years. A comparison between the PWV measured by the APEX WVR and the PWV from converted tipper opacity from tipper TA is shown in Figure~\ref{figure7}. The slope from the data, estimated using the procedure described in the previous section, is 1.011, which indicates a $1.1 \%$ difference between both measurements. The Pearson correlation coefficient for such dataset was calculated as described in~\cite{cortes16}, giving a value of $0.92$.

\begin{figure}[h]
\resizebox{\hsize}{!}{\includegraphics{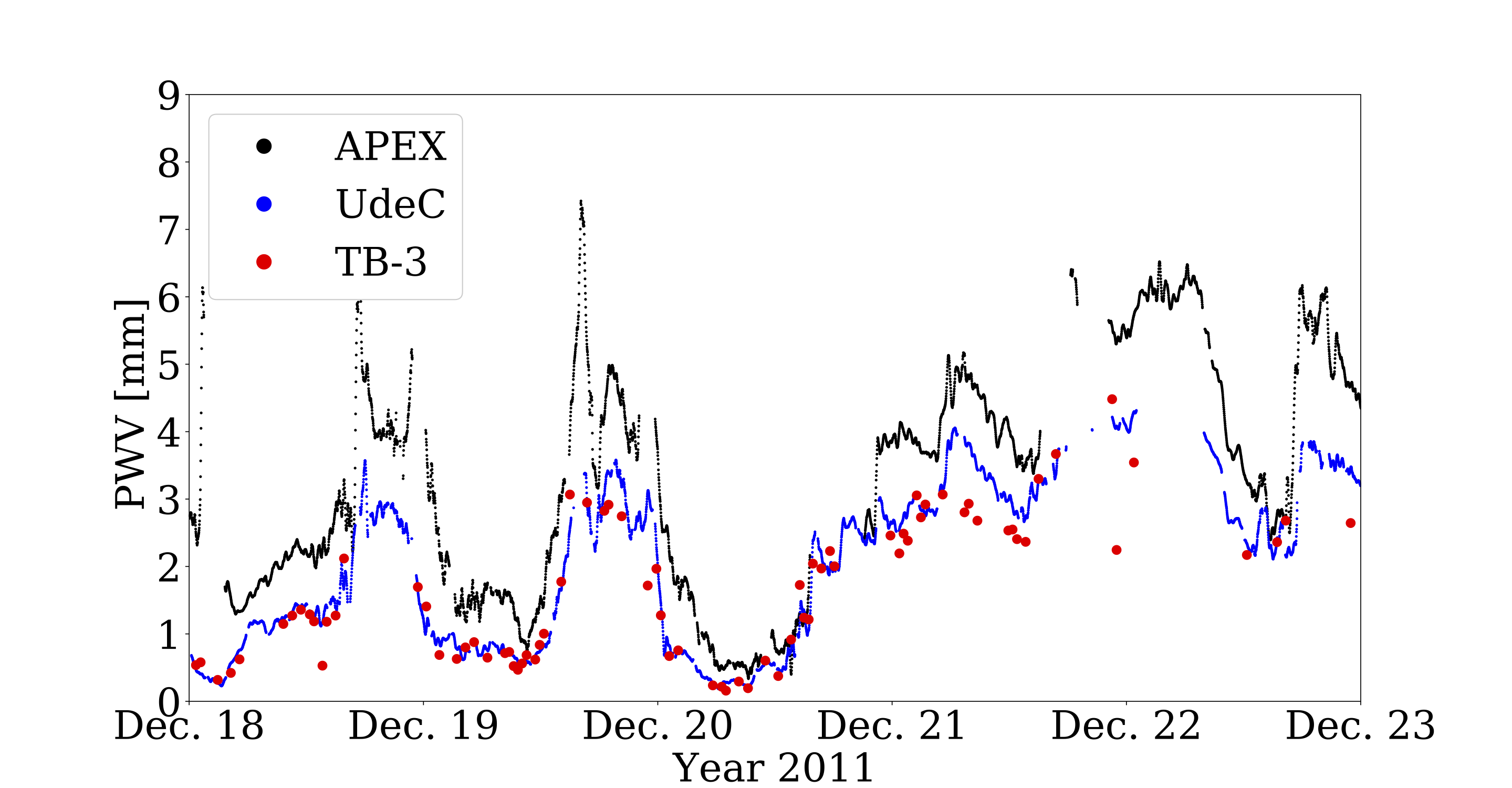}}
\caption{PWV reported for a period of five days from Cerro Chajnantor summit (data from UdeC-WVR and TB-3) and from the Chajnantor Plateau (data from the APEX WVR). The correlation between the UdeC and TB-3 data, both located at the Chajnantor summit, and the instantaneous difference with the PWV at the plateau measured by the WVR at APEX.}
\label{figure10}
\end{figure}

Figure \ref{figure10} shows a visual comparison between the results of the conversion method for the TB-3 tipping radiometer located at the summit of Cerro Chajnantor and data from the UdeC WVR, when concurrently deployed at the same site. The WVR data from the APEX observatory are reported as a reference and the PWV variations are well correlated for both sites. Interestingly and looking at Figure \ref{figure10}, we notice what could be the appearance of atmospheric inversion layers for certain periods of the data. These can be observed when the PWV at Cerro Chajnantor drops dramatically compared to the PWV at the plateau, which is also mentioned in \cite{bustos14}. The appreciable difference in absolute PWV on both sites is mainly due to the vertical distribution of water vapor, which drops monotonically as altitude increases.

We assessed the potential improvement of the revised methodology in data quality by calculating the standard deviation of the various procedures applied to the same dataset. This is referred to as the conversion error of the method and we compare the errors given by the old methodology against the errors from the revised method. The results show an improvement of $10\>\%$ with the new method as expected from the impact of the conversion curves shown in Figure~\ref{figure3}.

\section{Distribution of PWV\ from site comparisons}

In this analysis, we consider $20$ years of time-overlapped data from various instruments fielded at the Chajnantor area. Using the revised method presented in Section 2 to convert opacity to PWV, a recalculation of our quantitative comparison between the atmospheric conditions for sites of interest for deployment of millimeter and submillimeter astronomy instrumentation is performed. The results of the analysis are presented by comparing the atmospheric conditions for two sites and according to the procedure that was previously described in Section 4. As mentioned before, the two datasets are appropriately downsampled in time to bring them to the same cadence.

\begin{figure}[h]
\resizebox{\hsize}{!}{\includegraphics{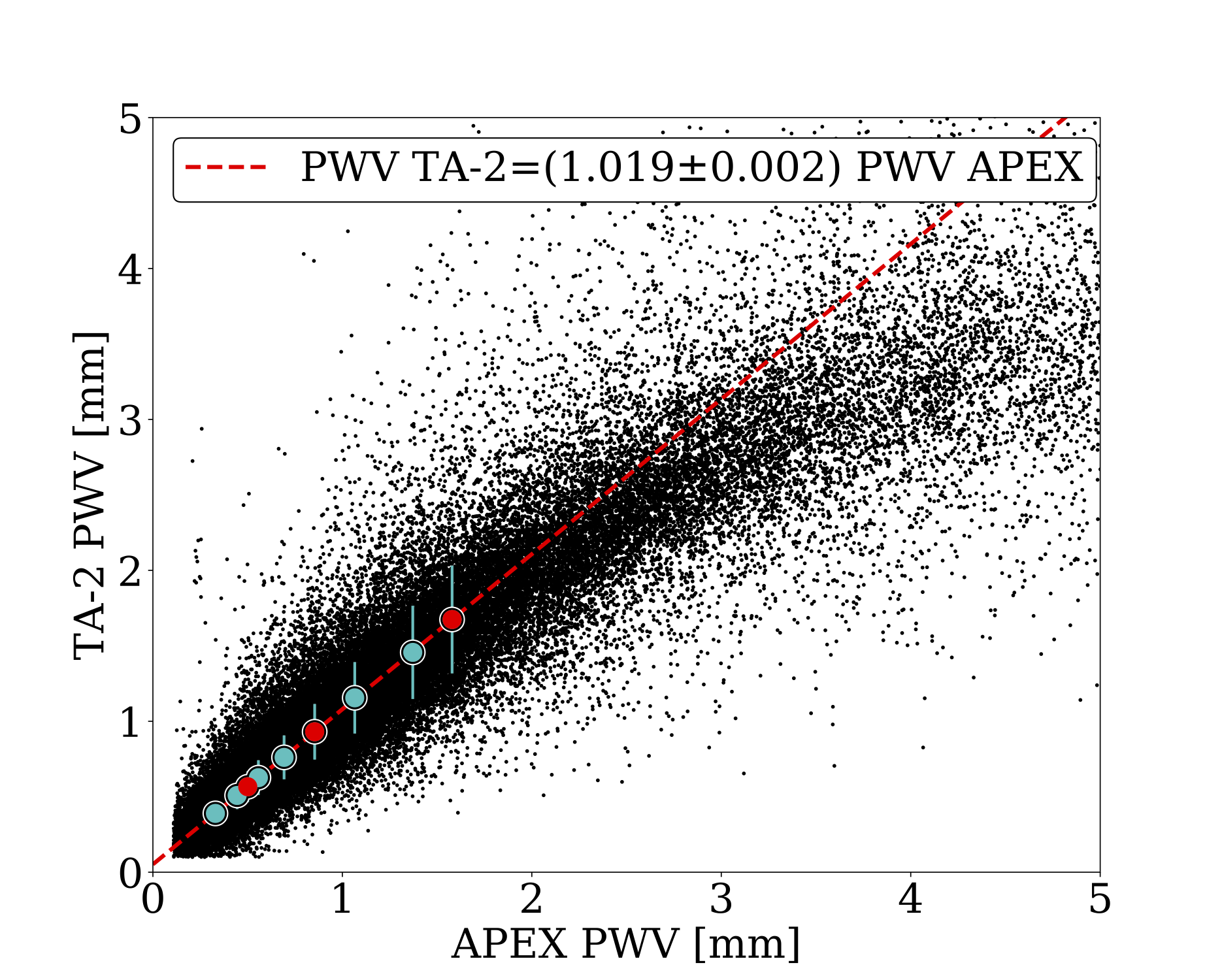}}
\caption{PWV determined from the APEX WVR vs. TA-2 converted PWV located at the CBI site, over the period of 2006 to 2010. The APEX site is $27$ m above the CBI site. The cyan circles are the PWV quantiles used in linear regression. The red circles indicate the quartiles ( 25\%, 50\%, and 75\%). Error bars are y-axis standard deviation for each quantile.}  
\label{figure9}
\end{figure}

The APEX VWR data versus TA-2 tipper converted PWV scatter plot is presented in Figure \ref{figure9}. The TA-2 was deployed at the CBI site; therefore it was $27$ m lower in altitude than the APEX site and about $2.5$ km away in linear distance. It is relevant to note that the CBI site is very close to the ALMA array center, only $720$ meters away; consequently, this site also faithfully represents the atmospheric characteristics for the submillimeter interferometer site. In Figure \ref{figure9}, the time-matched linear regression gives $1.9\%$ excess in PWV for the CBI site compared to the APEX site, which agrees with what is expected from the use of a standard atmosphere and the difference in their altitudes. The Pearson correlation coefficient for the linear regression in Figure \ref{figure9} is $0.86$.

\begin{figure}[h]
\resizebox{\hsize}{!}{\includegraphics{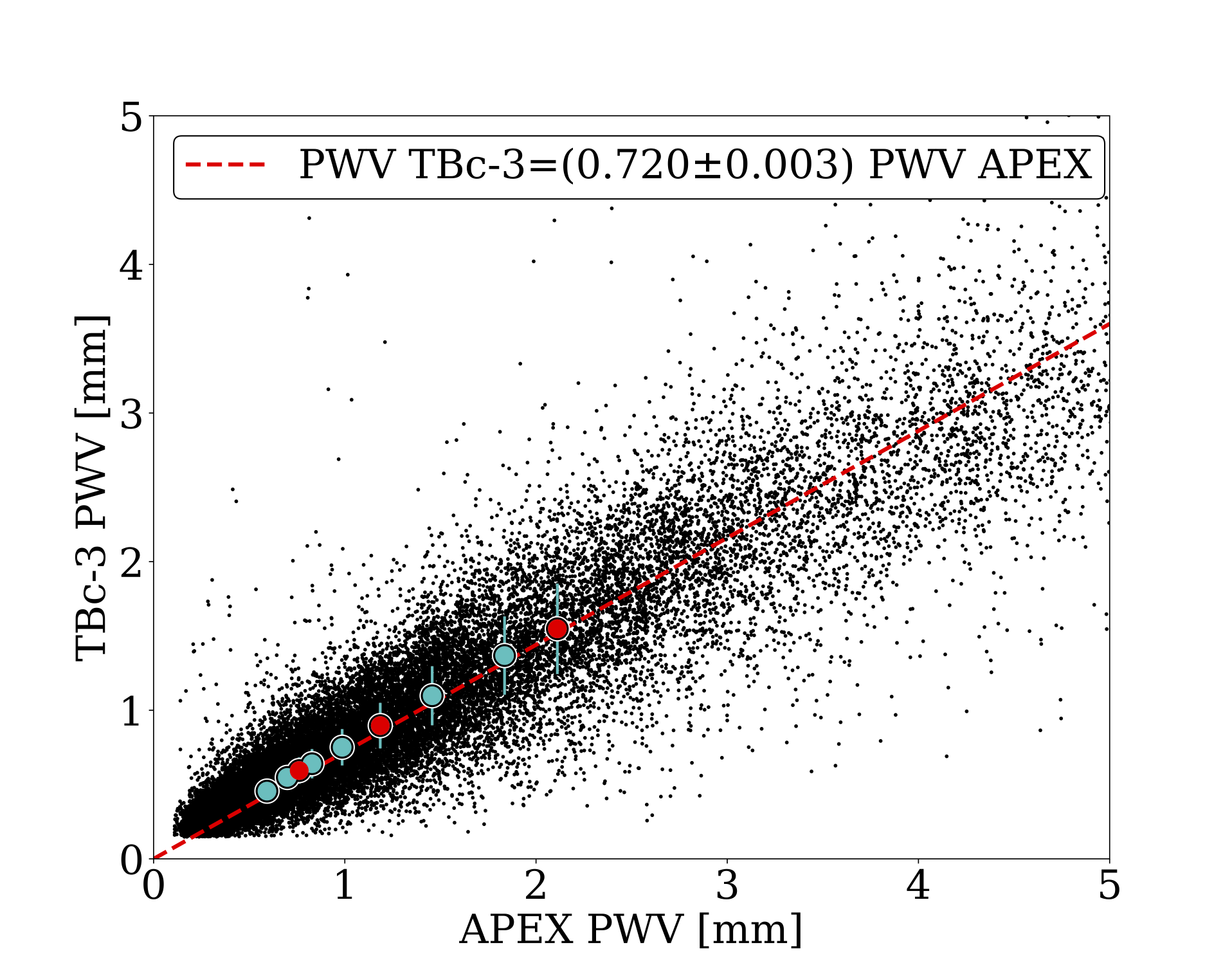}}
\caption{PWV comparison between Cerro Chajnantor and the plateau. Data from 2009 to 2012 are included in this figure. The cyan circles are the PWV quantiles used in linear regression. The red circles indicate the quartiles ( 25\%, 50\%, and 75\%). Error bars are y-axis standard deviation for each quantile. The slope in this graph denotes the significant drop in PWV at the Cerro Chajnantor summit as opposed to the Chajantor plateau, with a $505$~m difference.}
\label{figure8}
\end{figure}

Figure~\ref{figure8} shows PWV versus PWV scatter-plot results for APEX and TB-3. TB-3 is the third location for the submillimeter tipping radiometer at the summit of Cerro Chajnantor. We note that the location of TB-3 is $505$~m higher that the APEX site. In Figure \ref{figure8}, the slope of $0.72$ indicates an overall year-round reduction of $28\%$ in PWV when ascending in altitude from the Chajnantor Plateau to the Cerro Chajnantor summit. This result is consistent with the ratio of $0.7$ between these altitudes reported by \cite{otarola19}, using data from radiosondes launched from Antofagasta, and \cite{bustos14}, who report only five days of data.

\begin{figure}[h]
\resizebox{\hsize}{!}{\includegraphics{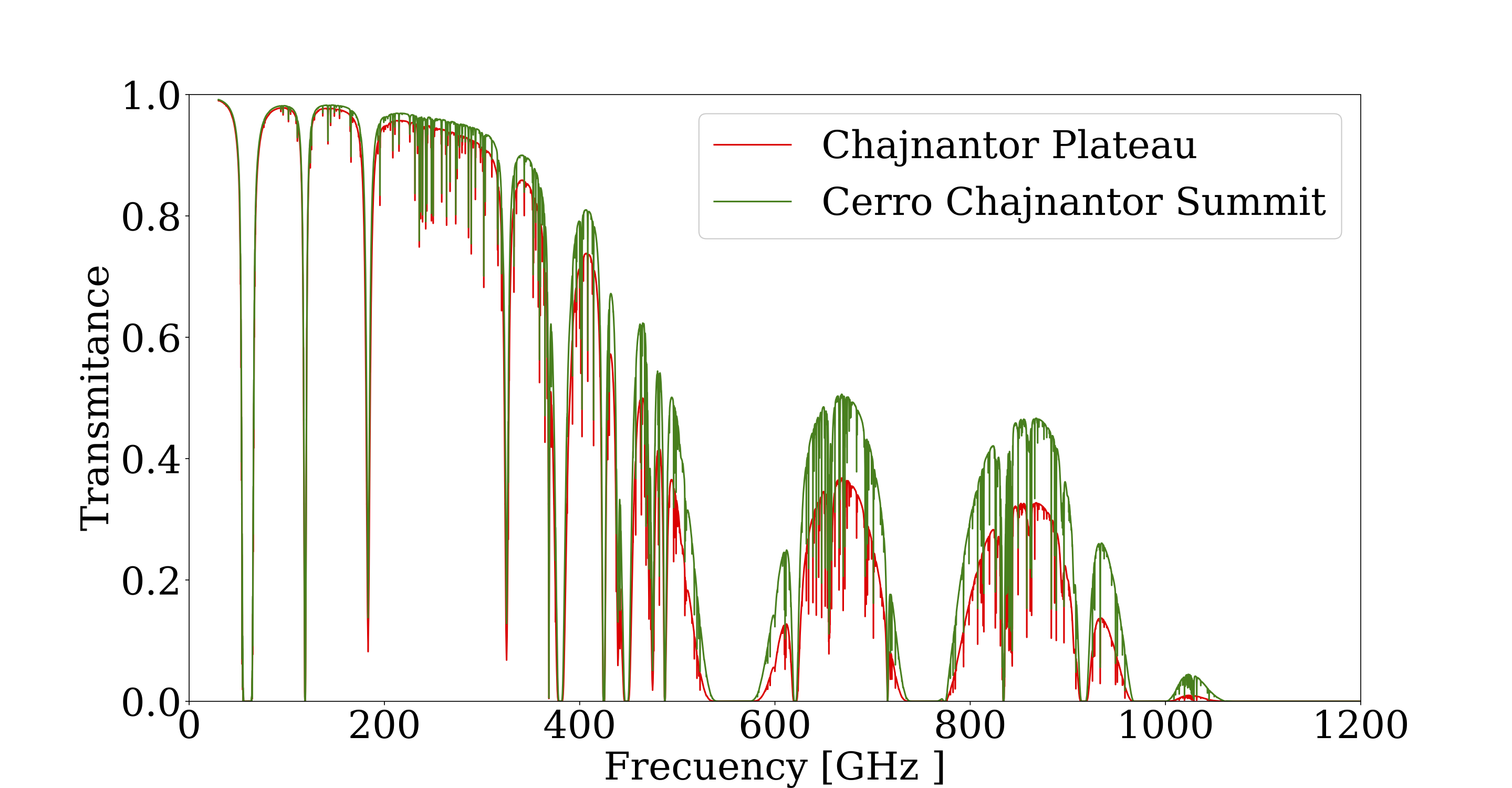}}
\caption{Atmospheric transmission for $30-1200$ GHz calculated with the AM software for the median measured conditions at the Chajnantor plateau and at the summit of Cerro Chajnantor.}
\label{figure34}
\end{figure}

The effect of the $28\%$ difference between the Chajnantor Plateau and Cerro Chajnantor summit on the atmospheric transmittance is shown in Figure \ref{figure34}. This image is the result of a simulation using the AM software for both sites. The input parameters in AM were $0^{\circ}$ of zenith angle, a frequency window from 30 GHz to 1200 GHz, a mean ground temperature of $272.6$ K, and $268$ K for Chajnantor Plateau and Cerro Chajnantor summit, respectively. The considered amount of PWV in the Chajnantor Plateau and for the Cerro Chajnantor summit was $1$ mm and $0.72$ mm, respectively. These values for PWV are taken from the 50th percentile result of our long-term analysis for both sites. As can be seen in Figure~\ref{figure34}, the impact in transmittance is more relevant at the higher frequency bands in the submillimeter regime and, thus, supports the installation of submillimeter astronomical equipment at the summit of Cerro Chajnantor. 

A summary with the results for the comparison between the characterized sites is presented in Table \ref{tabla-instr}. These results are calculated from PWV versus PWV derived slopes and by applying the proposed conversion method to the submillimeter tipper data. The instruments at the Chajnantor Plateau are APEX and TA-2 (CBI site). The instrument at Cerro Chajnantor summit is TB-3.

\begin{table*}[h]
\caption{\label{t7} Ratios of PWV between instruments at multiple sites in the Chajnantor area. The table includes time span of concurrent measurements, derived slopes between sites, the altitude difference between sites, and inferred scale height.}
\centering
\begin{tabular}{lcccc}
\hline\hline
      Instrument                    &     Period       &     Slope       &    Altitude & Scale  \\ 
   pairs               &                 (yrs)             & & diff  (m)& height (m)\\
\hline\\

 {\bf{ $\displaystyle\frac{\rm{TA-2}}{\rm{APEX}}$}}   &  {{ 06-10 }}        &  {\bf{ $1.019\pm0.002$ }}     &  -27 & $1434\pm149$ \\ \\
{\bf{ $\displaystyle\frac{\rm{TB-3}}{\rm{APEX }}$}}    &     {{09-12}}      &       {\bf{$0.720\pm0.003$}}    & 505& $1537\pm19$\\ \\
{\bf{$\displaystyle\frac{ \rm{TB-3}}{\rm{TA-2 }}$}}                     &  {{ 09-10}}        &  {\bf{$0.710\pm0.01$}}  &  532 &$1553\pm63$  \\ \\

\hline
\end{tabular}
\label{tabla-instr}
\end{table*}

The comparison between sites is consistent with the exponential decay in PWV of a standard atmosphere. This decay can be modeled as follows:

\begin{equation}
\hspace{30mm}\displaystyle PWV\>=\> PWV_{0}\cdot {\rm e}^{-\frac{\Delta h}{h_{o}}}
\label{eq1}
,\end{equation}

\noindent where $PWV_{0}$ is the PWV measured at the lowest altitude, $h_{0}$ is the scale height, and $\Delta h$ the difference in altitude for the two sites. Using the values reported in Table~\ref{tabla-instr}, we calculate a water vapor scale height for the Chajnantor Plateau of $1537$~meters. This scale height agrees with previous results from \cite{bustos00, giovanelli01, bustos14, cortes16, radford16, kuo17} 

\section{Twenty-year water vapor study for the Chajnantor Plateau}

In this section, we present the $20$ years of PWV data for the Chajnantor Plateau, from $1997$ to $2017$. We present our PWV dataset in full, gathering information from various types of instruments, but we convert these findings to the same unit. The APEX radiometer was included in this database for completeness. 

\begin{figure}[h]
\resizebox{\hsize}{!}{\includegraphics{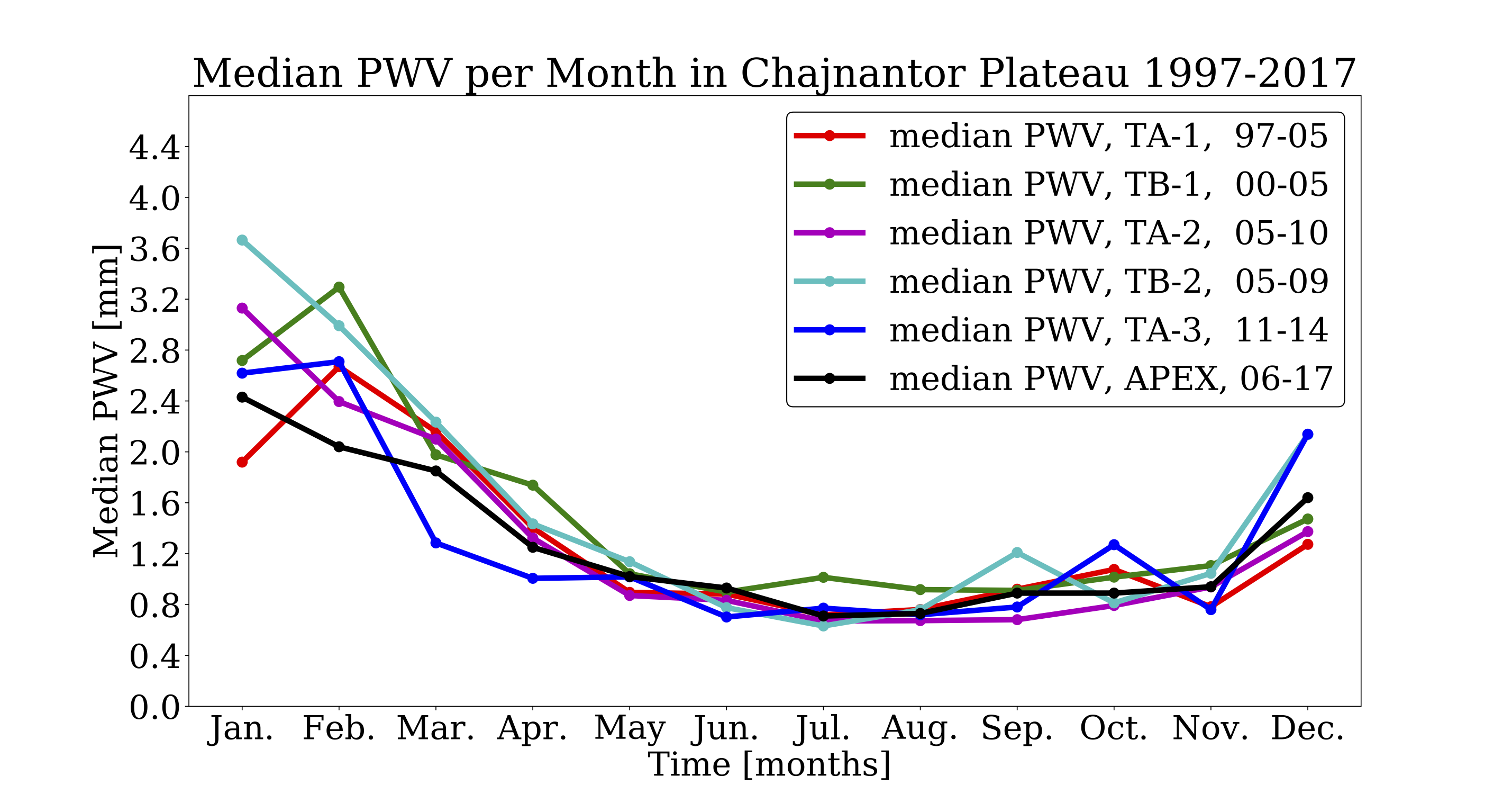}}

\caption{Median PWV per month per instrument located at the Chajnantor Plateau, over a $20$-year span. The highest value and dispersion for the mean PWV in the months of January and February and the lower PWV in the months of June, July, and August are noted. This plot shows the consistency between the measurements of the instruments.}
\label{figure19} 
\end{figure}

With the aim of understanding the general annual PWV cycles on a month-by-month basis at the Chajnantor Plateau, we present all the median PWV results per instrument in the period from $1997$ to $2017$ in Figure~\ref{figure19}. The PWV from APEX radiometer (with more cadence of data) is included in Figure~\ref{figure19} on purpose to compare the results per month versus the results of other instruments that have been converted to the same unit, providing consistency to the results from all instruments.

At the plateau, January and February are always shown as the wetter months of the year, while June through August, the dryest months, offer the best conditions for submillimeter astronomy observations. Taking the median of the data per month over the 20-year span of our study gives an overall view for the year-round climate at the Chajnantor site. Such analysis is shown in Figure~\ref{figure55}. We note again the drastic increase in PWV for the months December through March, which is a clear signature of local effect called Altiplanic Winter, which enhances the east-West water vapor circulation from the wet side of the Andes, north Argentina, Paraguay, and south Brazil into the west and through the Chajnantor area.

\begin{figure}[h]
\resizebox{\hsize}{!}{\includegraphics{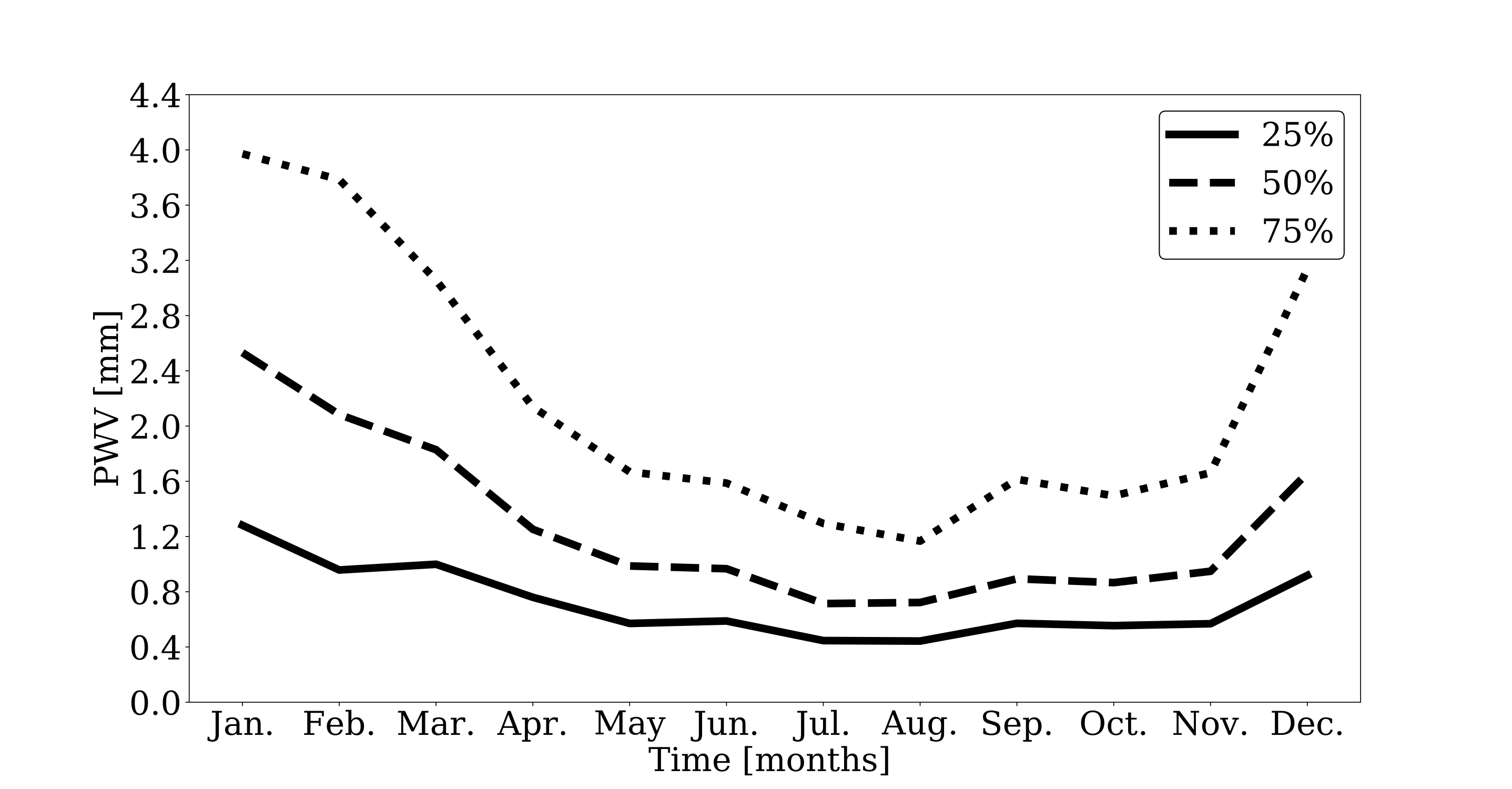}}\\
\resizebox{\hsize}{!}{\includegraphics{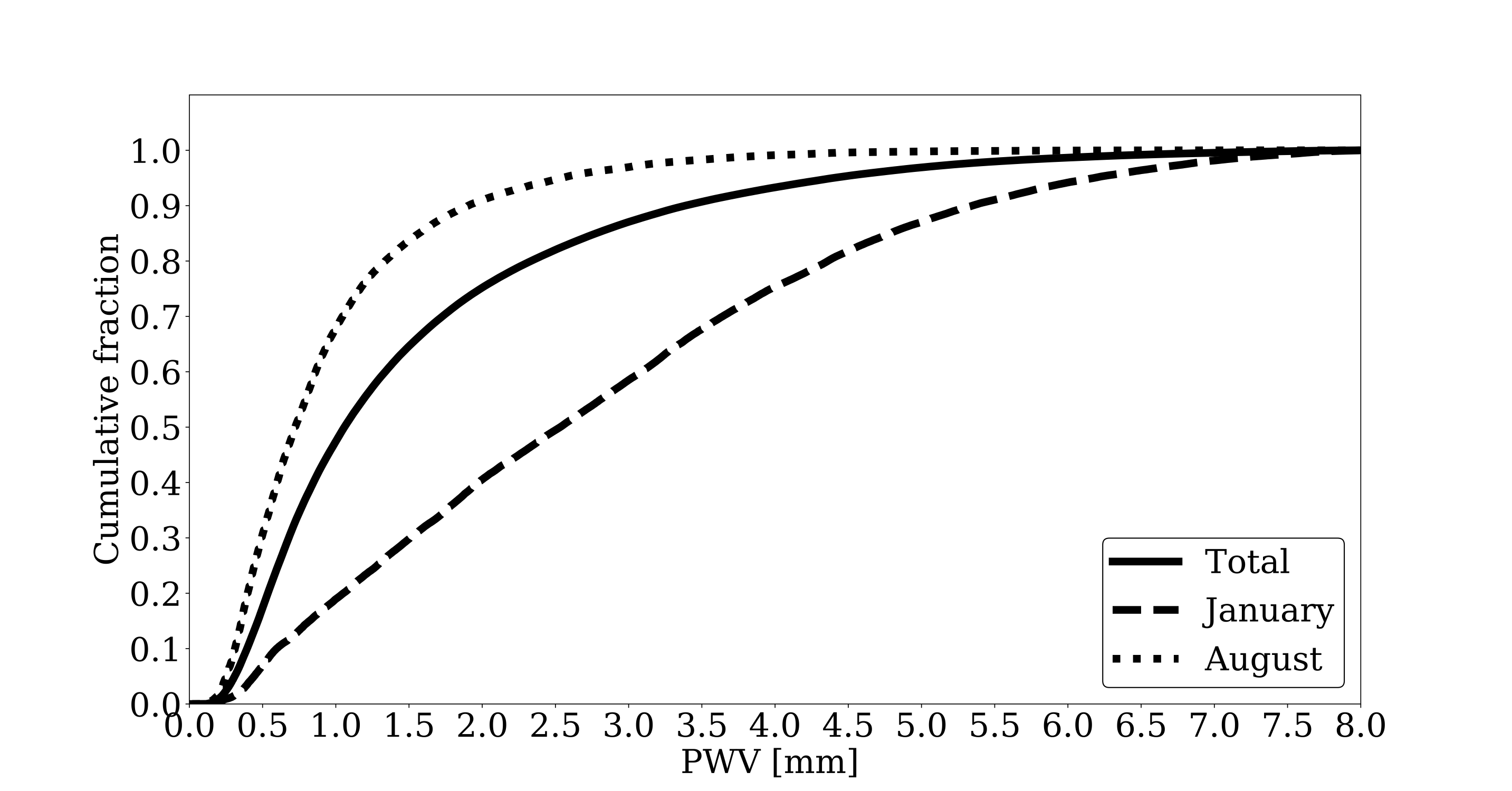}}\\
\caption{Twenty years of PWV study in Chajnantor Plateau. Monthly quartiles (25\%, 50\%, and 75\%) of PWV measurements over the full period (upper panel). On the plateau, the best observing conditions are shown over June, July, and August. Cumulative distribution fraction of PWV for Chajnantor Plateau is shown in the lower panel. Total fraction, along with the extreme cases of January and August, are included to limit the typical range of PWV for the remaining months.}
\label{figure30}
\end{figure}

The upper plot of Figure~\ref{figure30} is a plot of quartiles, which shows the variations over months of three different percentile distributions ($25\%$, $50\%,$ and $75\%$). The shape of the three curves is similar to Figure~\ref{figure19}. This figure allows a robust statistical assessment of the atmospheric quality for the Chajnantor Plateau. 
With the aim of understanding the most extreme months in the Chajnantor area, a cumulative fraction plot is presented at the bottom of Figure~\ref{figure30}. This is an assessment of the conditions for the Chajnantor Plateau (solid line), with the year-round, 50th percentile value near $1$~mm for the site. In addition, we find the median values for the extreme months of January and August, with a 50th percentile value for each condition of $2.56$~mm and $0.72$~mm, respectively.

\begin{table}[h]
\caption{\label{t7} Summary of PWV cumulative fractions for the Chajnantor Plateau. The extreme cases of January and August are also shown.}
\centering
\begin{tabular}{lccc}
\hline\hline
Period of time  &  $25\>\%$   &   $50\>\%$  &  $75\>\%$    \\ \hline
Year-round     &   0.60 mm       &   1.05 mm    &  1.98 mm  \\ 
January        &     1.28 mm     &  2.56 mm      &  3.97 mm \\ 
August     &   0.44 mm       &    0.72 mm    & 1.19 mm \\ 
\hline
\end{tabular}
\label{tab-acum}
\end{table}

The results from Figure \ref{figure19} are also consistent with the results from \cite{marin17}. These authors used the Climate Forecast System Reanalysis (CFSR) \citep{saha10} as source for the PWV data for Chajnantor and the APEX radiometer data ($2006$ to $2010$) to validate the analysis. In \cite{marin17}, the authors argued about a connection between the Madden-Julian Oscillation (MJO) pattern and the variability of the PWV at Chajnantor. Given that the MJO starts in the western Pacific Ocean and develops through the east sometimes reaching the South American coast, we believe a study of the MJO could be used as a tool to predict wet events in Chajnantor. Therefore, this study would be a good addition for the planning of scientific activities in the area.

\begin{figure}[h] 
\resizebox{\hsize}{!}{\includegraphics{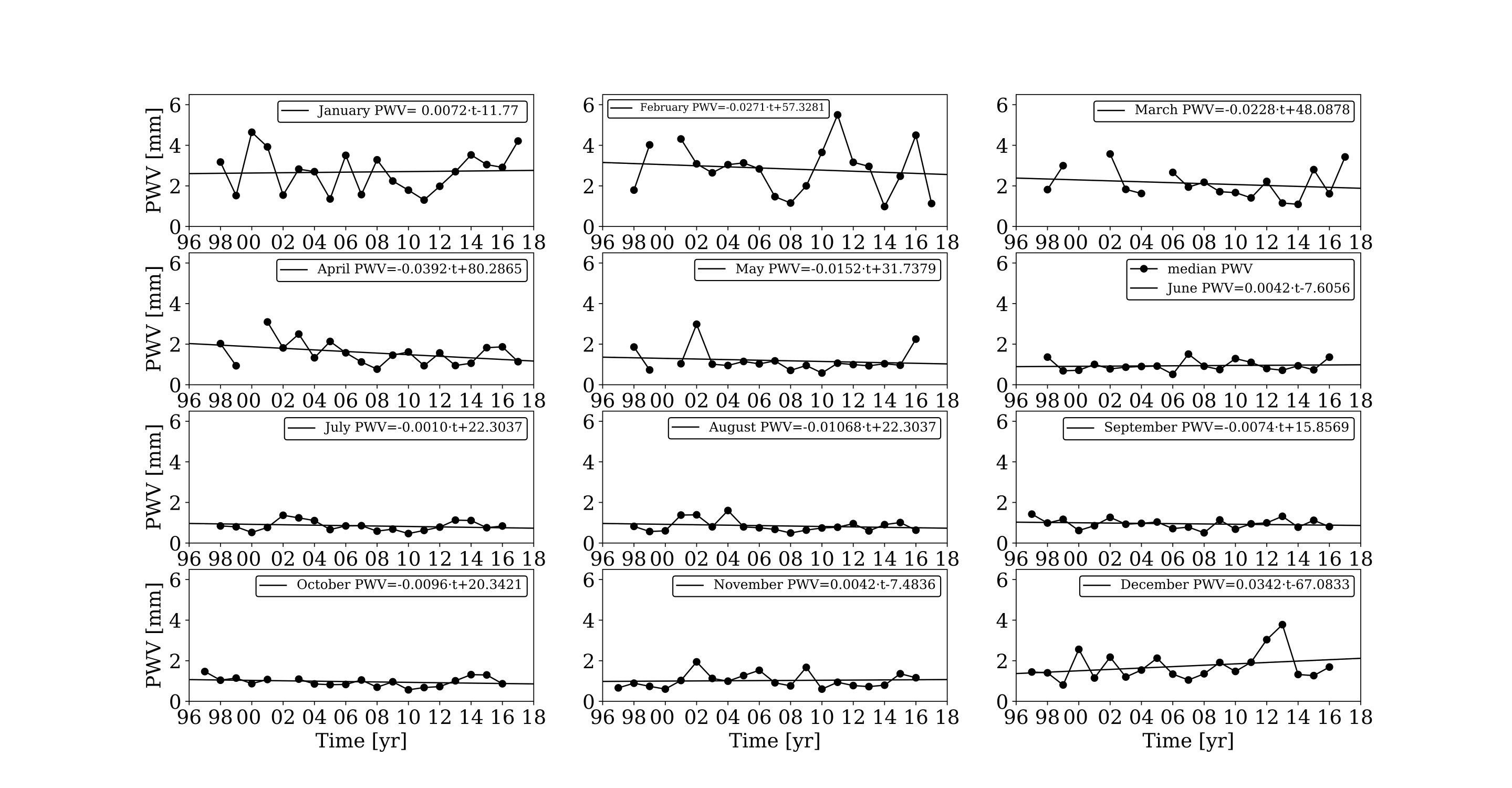}}
\caption{Median PWV per month over 1997-2017 in the Chajnantor Plateau. The dots indicate the median value of PWV in this year, while the straight line indicates the trend for the PWV over the period of the study, with a possibility of projecting future PWV values. The median for all the monthly slopes is $-0.001\pm0.014$, that is, consistent with zero. We do not see a long-term increase or decrease in the atmospheric variable of interest for Chajnantor.}
\label{figure20}
\end{figure}

The long-term median climatology of the Chajnantor Plateau is shown in Figures~ \ref{figure19} and ~\ref{figure30}. A possibility for studying large-scale anomalies in the climate for the Chajanantor area is to review the evolution of the PWV over the span of this study. We decided to assess such evolution by plotting the PWV for each month over the years in a single plot, as shown in Figure~\ref{figure20}. Each subplot in this figure corresponds to a single month and each data point in the plot corresponds to the median PWV from that month in our consolidated database. In addition, a linear regression  and their math expression is presented in each subplot. Each linear regression are in function of time (years) and with these expressions we can estimate the amount of PWV in the future for each month, given the data collected in this period. All the slopes are small; therefore, they do not indicate a clear tendency (increase or decrease) for the amount of PWV in the period of study or in the future for the Chajnantor Plateau.

\section{Cerro Chajnantor summit}

Cerro Chajnantor is one of the highest peak in the Chajnantor area. The atmospheric conditions at the peak are different from the Chajnantor Plateau, as expected given the altitude difference. In addition, the appearance of inversion layers affect the instantaneous value for PWV, drying out the summit of Cerro Chajnantor at a higher rate than the plateau.

\begin{figure}[h!]
\resizebox{\hsize}{!}{\includegraphics{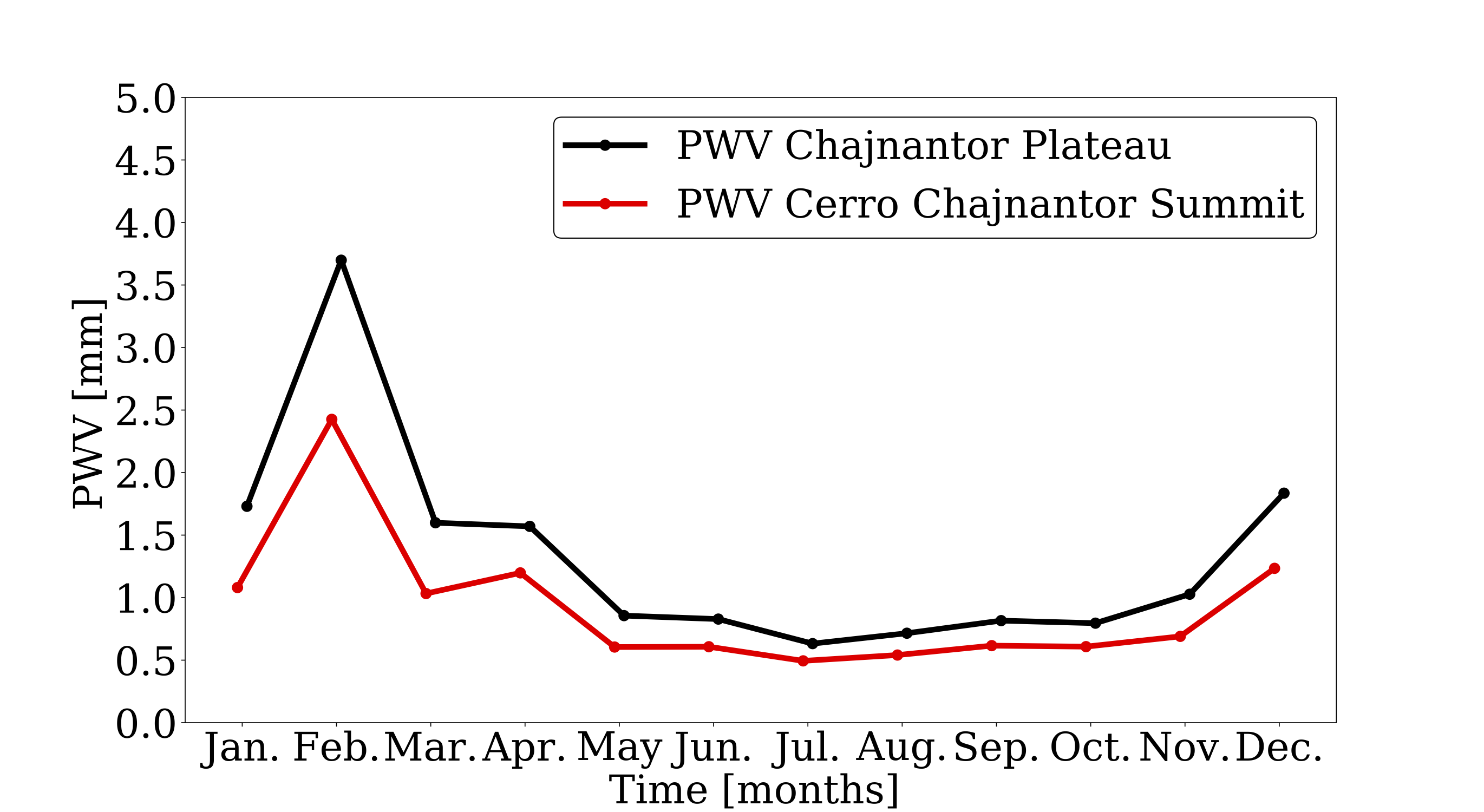}}
\caption{Comparison between Cerro Chajnantor Summit and Chajnantor Plateau.The differences per month in median values for PWV between both sites are shown. July and August show the least difference between both sites.}
\label{figure55}
\end{figure}

\begin{figure}[h!]
\resizebox{\hsize}{!}{\includegraphics{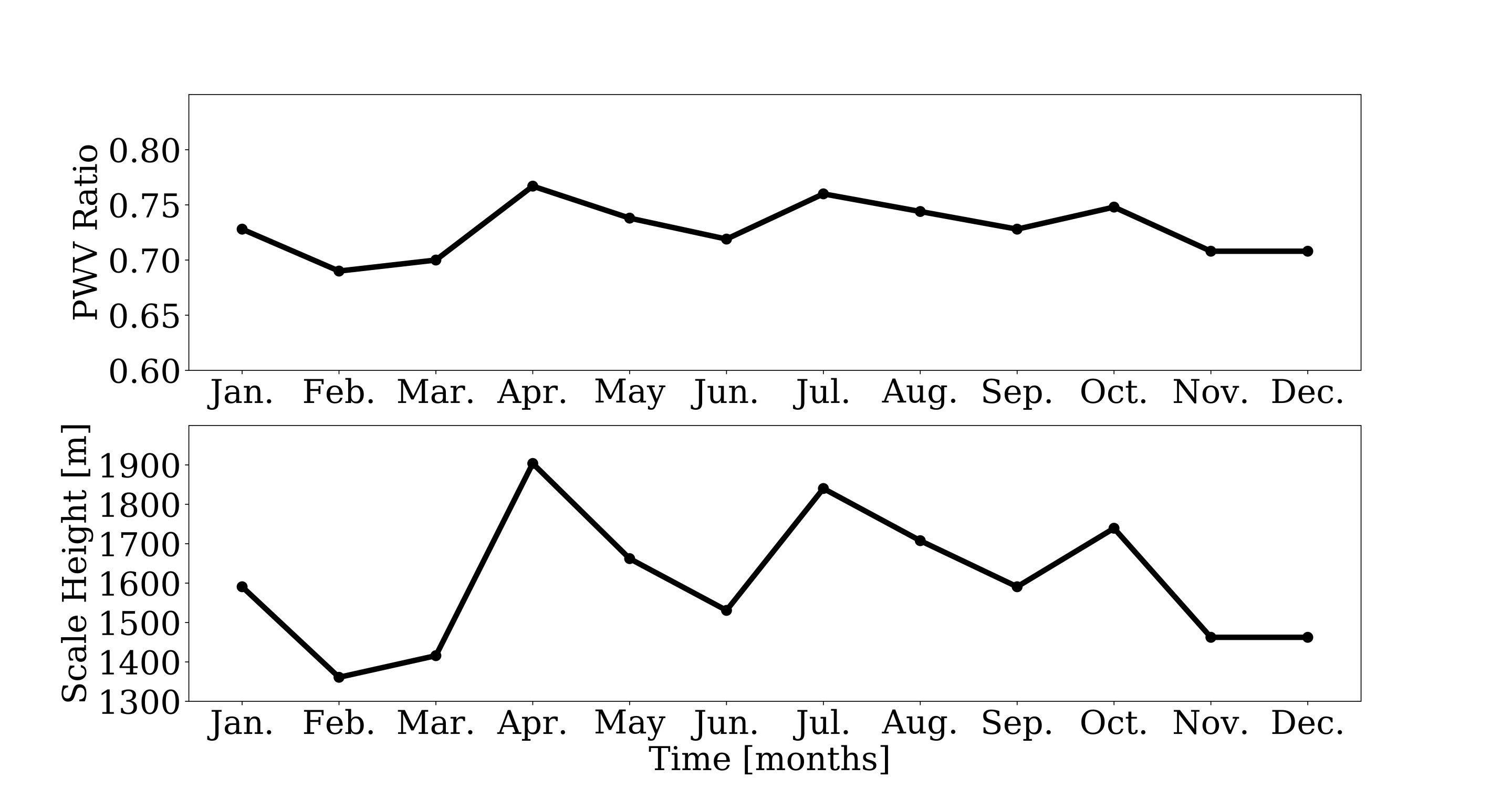}}
\caption{Chajnantor summit vs. plateau PWV ratio and scale heights. The upper plot shows the ratio $PWV_{cc}/PWV_{plateau}$ per month. The lower panel shows the scale height from the ratios obtained in the upper panel using the equation (1), considering $505$ m of altitude difference.}
\label{figure56}
\end{figure}

Median PWV values over the full span of our study are used to compare the atmospheric conditions between Cerro Chajnantor Summit and Chajnantor Plateau, as shown in Figure~\ref{figure55}. The Chajnantor Plateau always has greater values of PWV in comparison with the Cerro Chajnantor Summit. The months of December, January, and February show the highest values of PWV and error-bar size. February is considered a special case since less data are available for this month and the data are highly variable. High PWV (bad weather) drives the instruments into nonlinear conditions and turns them off when safety conditions are triggered. This might explain the anomalous values of PWV for that month. Ratios of PWV with data from both sites are presented in the upper panel of Figure~\ref{figure56}. Equation 1 is used to calculate the water vapor scale height for the Chajnantor area, using an altitude difference of $505$ m. As expected, the water vapor scale height is highly time variable. 

\begin{figure}[h]
\resizebox{\hsize}{!}{\includegraphics{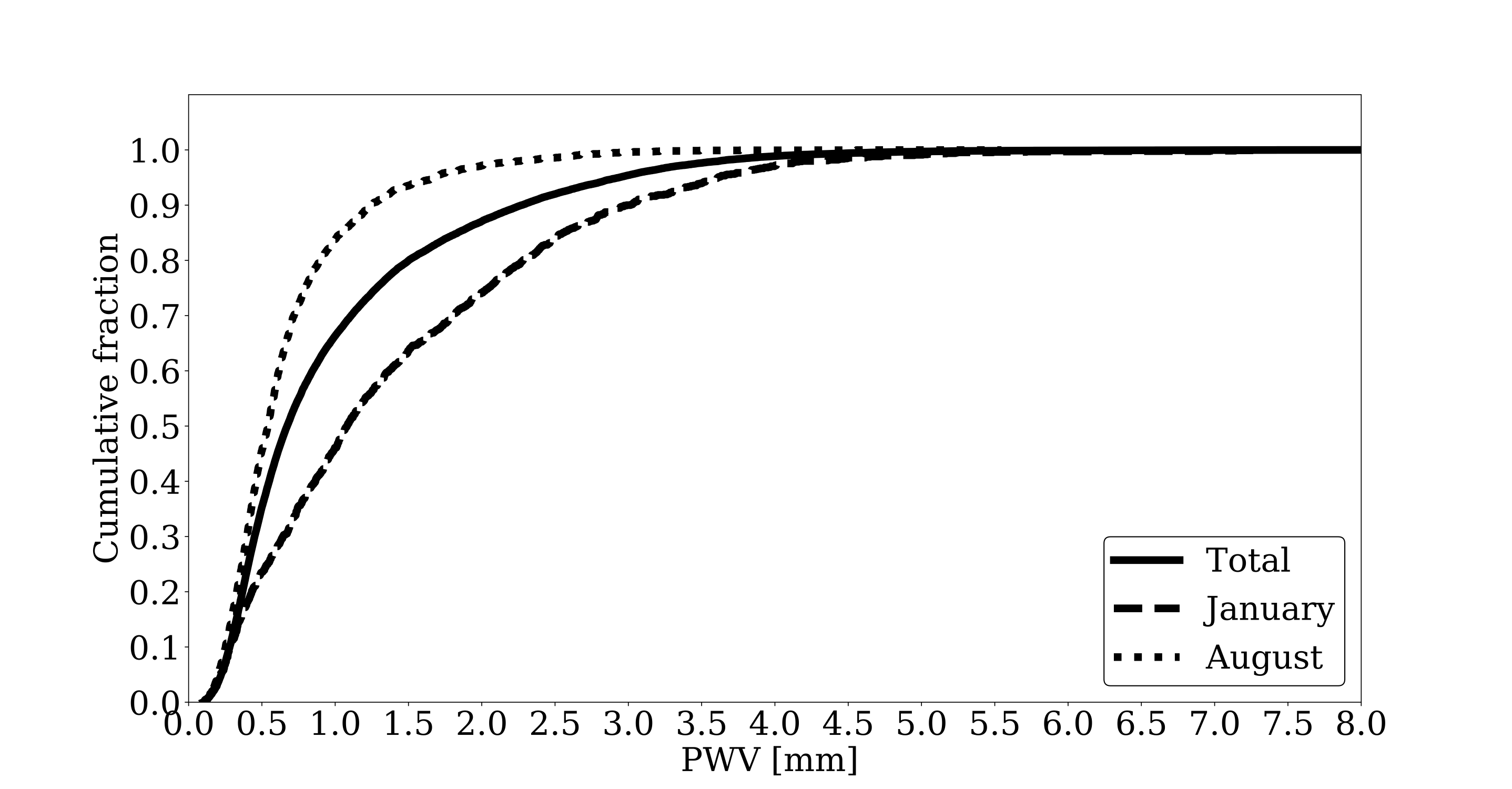}}
\resizebox{\hsize}{!}{\includegraphics{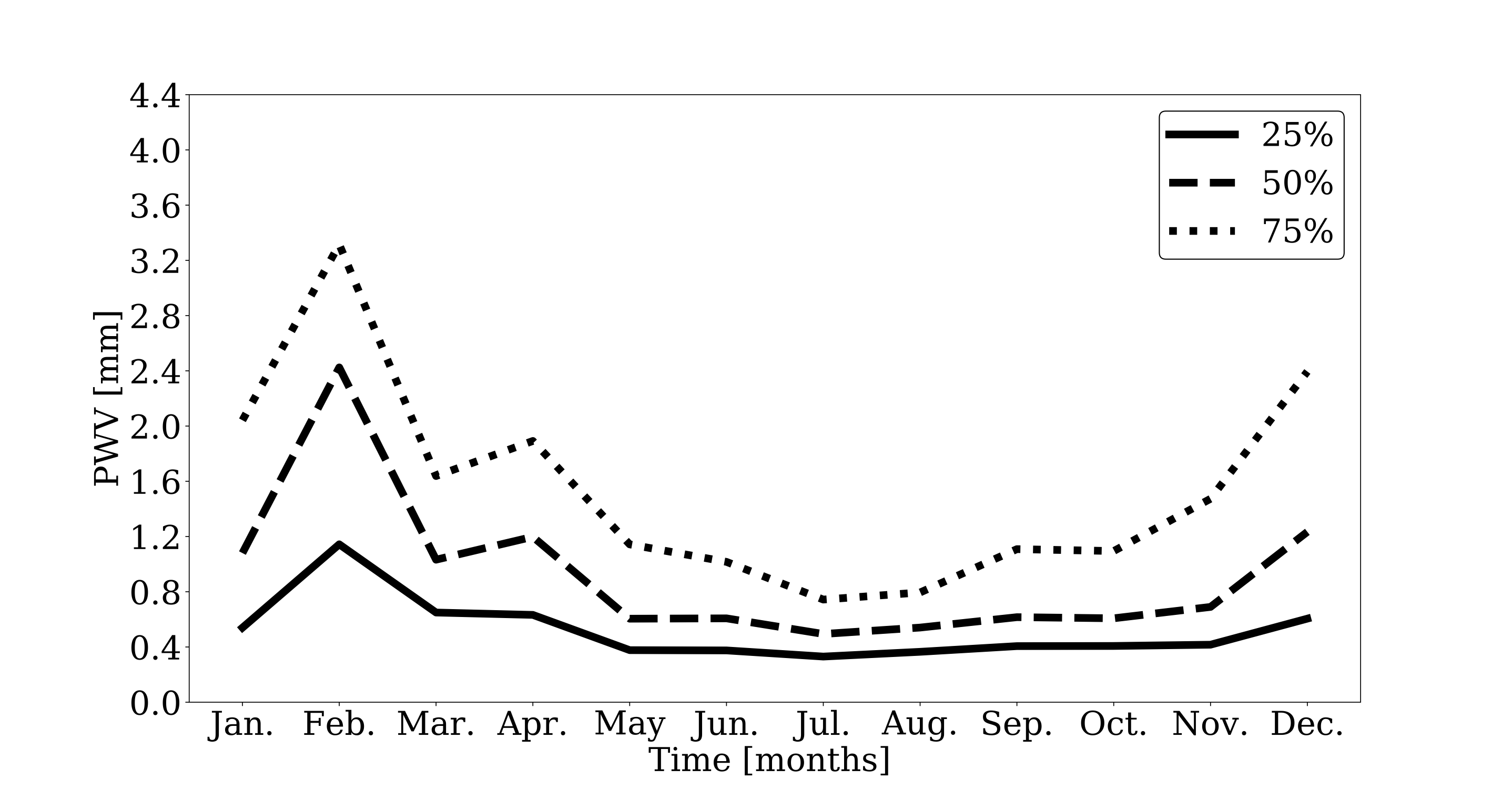}}
\caption{PWV data for the summit of Cerro Chajnantor. Five years of study are included in this analysis. The upper panel shows the cumulative distribution fraction of PWV. The lower panel shows the monthly quartiles (25\%, 50\%, and 75\%) of the PWV measurements for the summit.}
\label{figure303}
\end{figure}

Cumulative fraction plots for the Chajnantor summit are shown in Figure~\ref{figure303}, confirming the very good submillimeter observing conditions at the site. In summary, it offers less than $0.5$ mm for at least $30\%$ of the time, during the austral winter, which is outstanding for ground-based submillimeter access. Again, we hope this information can be appropriately used for the observations planning by all the instruments located into the area. The PWV quartiles were extracted from the lower panel of Figure~\ref{figure303} and are presented in Table \ref{tab-acum}. The reported quartiles from Table \ref{tab-acum} are in good agreement with the results recently shown in \cite{radford16, otarola19}.\\

\begin{table}[h]
\caption{\label{t7} Summary of PWV cumulative fractions for the summit of Cerro Chajnantor summit. The extreme cases of January and August are shown and are added for reference.}
\centering
\begin{tabular}{lccc}
\hline\hline
Period of time  &  $25\>\%$   &   $50\>\%$  &  $75\>\%$    \\ \hline
Year-round     &  0.36 mm     &   0.67 mm   & 1.28 mm  \\ 
January       & 0.53 mm   &  1.08 mm    &  2.04 mm \\ 
August    &   0.36 mm       &    0.54 mm    &  0.79 mm\\ 
\hline
\end{tabular}
\label{tab-acum}
\end{table}

\section{Conclusions}
In this paper, we present the assembly of a $20$-year long database for the PWV conditions at the Chajnantor area, covering from $1997$ to $2017$. Multiple instruments with different variables were used to compile such a large dataset. However, they were appropriately calibrated and converted to make the largest possible climatic evaluation of the site. The results of this study can be used, in addition to the scientific value, to plan observatory operations. The millimeter and submillimeter telescopes, such as ALMA, APEX, Atacama Submillimeter Telescope Experiment (ASTE), NANTEN2 and the future CCAT-prime, and LCT, will be able to plan the use of instruments at specific wavelengths depending on the month, that is, a 350 $\mu$m or near terahertz instrument should only be available during austral winter months.

The methodology to convert atmospheric opacity now includes the use of the instantaneous ground temperature as an input parameter, which contributed to produce more accurate results, as seen on Figure~\ref{figure3}. The cadences for the various instruments were also matched in order to appropriately use the data for a comparative analysis of the various sites. It was found that more robust and statistically significant results, for PWV ratios between different sites and instruments, are calculated if the PWV versus PWV scatter plots with time-matched samples are replaced with cumulative fractions for each site or instrument.

Regarding the comparison between the Chajnantor Plateau and the summit of Cerro Chajnantor, we found a decrease in PWV for the summit versus the plateau of $28\%$, confirming the summit as a great submillimeter astronomy site. This result is in good agreement with previous works  \citep{bustos14, radford16, cortes16, otarola19}. In addition, we found differences in the PWV within the plateau  (i.e., the CBI site), which is $27$~m below the APEX site, shows a $1.9\%$ excess in PWV compared to the APEX site, which is consistent with a model of standard atmosphere. We calculated a year-round atmospheric scale height for the Chajnantor area of $1537$~m. This scale height also agrees with previous works from \cite{cortes16, radford16, kuo17}, whom reported a scale height of $1475$~m for the same area. 

Given the conversion method that has been presented, the tipper radiometers are validated and can be used to characterize other sites of interest for the installation of future telescopes and their logistical considerations. Using the appropriate atmospheric model for the site under measurements, the system does not require external calibrators to deliver PWV, as has been assumed in the past. 

Interestingly, our long-term study of the PWV conditions at the Chajnantor Plateau did not show evidence of PWV trends, neither an increase nor a decrease over the 20 years of evaluation.

The AM configuration files used in this paper can be requested by e-mail from F. Cort\'es (fercortes@udec.cl).

\section*{Acknowledgements} 

We thank the people and observatories who contributed to this project. This publication is based on data acquired with the Atacama Pathfinder Experiment (APEX). APEX is a collaboration between the Max-Planck-Institut fur Radioastronomie, the European Southern Observatory, and the Onsala Space Observatory.

R. Reeves and F. Cort\'es acknowledge support from Fondo Gemini-Conicyt programa de astronom\'ia/pci folio $32140030$. R. Reeves acknowledges support from CONICYT project Basal AFB-170002 and from Fondo Astronom\'ia QUIMAL-CONICYT 140005.

K. Cort\'es acknowledges support from Conicyt with her PhD fellowship.

R. Bustos acknowledges support from UCSC internal funds FAA Nº 259/2019 and Fondo Astronom\'ia QUIMAL-CONICYT 180003.

R. Reeves acknowledges support from CONICYT project Basal AFB-170002 and from Fondo Astronomía QUIMAL- CONICYT 140005 / QUIMAL-CONICYT 160012.

\small{
\bibliographystyle{cys}
\bibliography{biblio}
\bibliography{biblio}
}

\end{document}